\documentclass[a4paper,11pt,oneside]{book} 
\usepackage{amssymb, bm, amsmath}
\begin{document}
\begin{titlepage}
\thispagestyle{empty}

\hoffset 0.25in
\begin{center}
\vskip 0.5cm
{\LARGE{\emph{Dynamical aspects of jovian irregular satellites}}}\\
\vskip 4.0cm
{\large{A thesis submitted for the degree of}}
\vskip 0.5cm
{\large{Doctor of Philosophy}}
\vskip 2.0cm
{\large{by}}
\vskip 1.0cm
{\large{Tobias Cornelius Hinse, B.Sc., M.Sc.}}
\vskip 2.0cm
{\large Armagh Observatory} \\
{\large Armagh, Northern Ireland, UK} \\
\vskip 1.0cm
{\large \&} \\
\vskip 1.0cm
{\large Faculty of Engineering and Physical Sciences}\\
{\large School of Mathematics and Physics} \\
{\large Queens University Belfast} \\
{\large Belfast, Northern Ireland, UK}
\vskip 1.0 cm
{\large{August 2010}}
\end{center}
\end{titlepage}

\thispagestyle{empty}

\begin{center}
\Large \emph{Publications}
\end{center}

This thesis resulted in the following publications:

\begin{itemize}

\item{Hinse, T.~C., Christou, A.~A. ``\emph{Mapping Phase Space Topology Structure and Dynamics of Jovian Irregular Satellites}'', 2008. Poster presentation (2008 DPS meeting at Cornell University, Ithaca, USA). 
AAS/BAAS, vol.~40, p.~481}

\item{Hinse, T.~C., Christou, A.~A., Alvarellos, J.~L.~A. and Go{\'z}dziewski, K. ``\emph{Application of the MEGNO technique to the dynamics of Jovian 
irregular satellites}'', 2010. MNRAS, vol.~404, p.~837.}

\end{itemize}

\thispagestyle{empty}
\noindent
This thesis is submitted for evaluation in accordance with the requirements 
for obtainment of the Degree of Doctor of Philosophy (PhD) of the Queens 
University Belfast, Northern Ireland, UK. I certify that the thesis presented 
by me for examination of the PhD degree is solely my own work other than where 
I have clearly indicated that it is the work of others.
\noindent
\vskip 1.0cm

\noindent
{\large{\bf External examiner:}}\\
\noindent
Prof. Dr. Carl D. Murray \\
School of Mathematical Sciences \\
Queen Mary, University of London \\
London, UK
\vskip 1.0cm

\noindent
{\large {\bf Internal examiner:}} \\
Prof. Dr. Mark E. Bailey\\
Armagh Observatory\\
Armagh, UK
\vskip 1.0cm

\noindent
{\large{\bf Thesis supervisor:}}\\
Dr. Apostolos Christou \\
Armagh Observatory \\
Armagh, NI,UK\\

\vskip 2.0cm

\noindent
{\underline{~~~~~~~~~~~~~~~~~~~~~~~~~~~~~~~~~~~~~~~~~~~~~~~~}}\\
\noindent
Tobias Cornelius Hinse, B.Sc, M.Sc. \\
Armagh Observatory \\
College Hill\\
BT61 9DG, Armagh\\
Northern Ireland, UK\\
(tobiash@arm.ac.uk, tchinse@gmail.com)

%%%%%%%%%%%%%%%%%%%%%%%%%%%%%%%%%%%%%%%%%%%%%%%%%%%%%%%%%%%%%%%%%%%%%%%%%%%%%%%%
%				ABSTRACT PART                                  %
%%%%%%%%%%%%%%%%%%%%%%%%%%%%%%%%%%%%%%%%%%%%%%%%%%%%%%%%%%%%%%%%%%%%%%%%%%%%%%%%

\begin{center}
\section*{Abstract}
\end{center}

This thesis concerns the mapping of chaotic resonances and the long-term dynamics of jovian irregular satellites. In order to obtain a detailed dynamical picture of the phase space structure occupied by the observed satellites we applied the numerical MEGNO (Mean Exponential Growth Factor of Nearby Orbits) technique to quantitatively detect chaotic resonant dynamics. By numerically following an unprecedented ensemble of test satellites we succesfully 
identified the location of orbital resonances and their occupation in phase
space. We carried out a complete mapping of the chaotic topology of satellite phase space in the form of high resolution MEGNO maps. In order to associate orbital resonances with their respective dynamical effects we considered solar and saturnian perturbations separately. In the restricted three-body (Jupiter-satellite-Sun) problem we show that \emph{the phase space occupied by retrograde jovian irregular satellites is dominated by numerous solar high-
order mean motion resonances}. These resonances are characterised by showing dynamical properties associated with chaos. The MEGNO technique also allowed us to detect the location of the secular resonance ($\varpi - \varpi_{\odot} \simeq 0$) when including Saturn's perturbing effects. Furthermore, the orbits of the satellites Carpo (prograde) and S/2003 J02 (retrograde) are found to be close to chaotic regions. Using single-orbit integrations we obtained numerical evidence that S/2003 J02 possibly exhibits long-term stable (or ``sticky'') chaos.

The location of solar mean motion resonances curiously coincides with satellite members of the retrograde Pasiphae family exhibiting a significant orbital dispersion in $(a, e)$ and $(a, I)$ space. \emph{Based on this result, we considered the hypothesis that long-term orbital dispersion is driven by solar high-order chaotic mean motion resonances}. Assuming that retrograde satellite families originated from a single collisional break-up event the process of chaotic diffusion by mean motion resonances could provide an effective transport mechanism in phase space, possibly solving a long-lasting conundrum as pointed out by (Nesvorn{\'y} et al., 2003, AJ, 126, 398) and (Nesvorny{\'y} et al., 2004, AJ, 127, 1768). Using Gauss's equations we calculated the observed velocity dispersion of the retrograde jovian satellite families to be of order $\delta V \simeq 320~ \textnormal{ms}^{-1}$. This is significantly larger than expected from the kinematics of a collisional break-up event. Numerical hydrocode simulations and laboratory impact experiments suggest a typical velocity dispersion on the order of a few tens of $\textnormal{ms}^{-1}$. 

To test our hypothesis we carried out long-term numerical orbit integrations using accurate adaptive time step and fast symplectic algorithms. For each retrograde satellite family we adopted an isotropic ejection model to generate initial conditions of test particles representing the initial state of a post-collisional fragmentation cloud. The particles in each fragmentation cloud were centred around the most massive (largest) satellite and we nu merically integrated the system over 4 - 5 Gyrs considering only solar perturbations. As a result we were able \emph{to demonstrate insignificant chaotic orbital diffusion in proper element space of retrograde satellites by solar high-order mean motion resonances}. 

In another attempt to identify the underlying dynamical mechanism capable of increasing the velocity dispersion of satellite fragments, produced by a collisional break-up event, we studied the effects of long-term perturbations by Saturn. We find that long-term chaotic diffusion in eccentricity and inclination is strongly associated with secular perturbations involving exchange of angular momentum between the satellite orbit and Saturn. Our results could partially reproduce the observed distribution of retrograde orbital mean elements. This finding supports our initial assumption of a collisional break-up event for all three retrograde satellite families.

Finally, due to chaotic diffusion of the proper inclination we numerically demonstrate the possibility of contamination of the Carme family with fragment members originating from the Pasiphae family. Observational support from photometric surveys indicate the existence of colour differences among Carme members suggesting contamination of the Carme family, assuming a homogenous progenitor satellite. We propose further photometric follow-up observations in order to test and further constrain these ideas. 

In summary, we have provided the first detailed mapping of jovian irregular satellite phase space using MEGNO; we have investigated a dynamical explanation for the relatively large velocity dispersion of the identified families of jovian irregular satellites, and have shown that their origin is consistent with formation in a primordial break-up event. The resulting fragments then experienced a subsequent dynamical diffusion of orbital elements primarily driven by secular perturbations of Saturn.
%%%%%%%%%%%%%%%%%%%%%%%%%%%%%%%%%%%%%%%%%%%%%%%%%%%%%%%%%%%%%%%%%%%%%%%%%%%%%%%%

\end{document}